\documentclass{aa}

\usepackage{graphicx}

\def\cds{$^{12}$C$^{16}$O }
\def\cts{$^{13}$C$^{16}$O }
\def\ctd{$^{13}$C$^{18}$O }
\def\cdsp{$^{12}$C$^{16}$O}
\def\ctsp{$^{13}$C$^{16}$O}
\def\ctdp{$^{13}$C$^{18}$O}
\def\cij{C$_{ij}$(J) }

\def\esspb{{\it E(6)}$^1$$\Pi$ }

\def\qpsz{4{\it p$\sigma$}(0)}
\def\qpszb{4{\it p$\sigma$}(0) }
\def\spi{$^1$$\Pi$}
\def\spib{$^1$$\Pi$ }

\def\sspb{$^1$$\Sigma$$^+$ }

\def\qppz{4{\it p$\pi$}(0)}
\def\qppzb{4{\it p$\pi$}(0) }

\def\tdpub{3{\it d$\pi$}(1) }

\def\tppzb{3{\it p$\pi$}(0) }

\def\xpzqp{X$^+$(0)4{\it p} } 
\def\xputdp{X$^+$(1)3{\it d$\pi$} }

\def\tppxz{3p$\pi$$^1$$\Pi$(E)-X$^1$$\Sigma$$^+$(0-0) }

\def\xsmix{X$^+$(0)4p$\sigma$ }

\def\AAb{{\AA} }
\def\mAAb{m{\AA} }
\def\mAA{m\AA}

\begin{document}

\title{Oscillator strengths for transitions to Rydberg levels in
 \cdsp, \cts and \ctd between 967 and 972 \AA\thanks{Based on experiments 
done at the Super-ACO electron storage ring at Orsay (LURE), Fr}}

  \author{M. Eidelsberg\inst{1},
 J.L. Lemaire\inst{1,2}, J.H. Fillion\inst{1,2}, 
 F. Rostas\inst{1}, S. Federman\inst{3} and Y. Sheffer\inst{3}}

\offprints{michele.eidelsberg@obspm.fr}   

\institute{
LERMA, UMR 8112 du CNRS, Observatoire de Paris, 92195 Meudon Cedex, France 
\and 
LERMA/LAMAp, UMR 8112 du CNRS, Universit\'e de Cergy-Pontoise, 95031 Cergy-Pontoise, France.
\and 
Department of Physics and Astronomy, University of Toledo, Toledo, OH53190 USA. 
}
   \date{Received 05/04/2004; accepted 11/05/2004}
\abstract{
Absorption oscillator strengths have been determined from high-resolution spectra in the 967-972 \AAb region 
of three CO isotopomers for transitions to the Rydberg levels \qppz, \tdpub and \qpsz, as well as to the mixed
{\it E(6)} level recently characterized by Eidelsberg et al. (2004). Synchrotron radiation from the Super-ACO
electron storage ring at Orsay (LURE) was used as a light source. Oscillator strengths were extracted from the 
recorded spectra by least-squares fitting of the experimental profiles with synthetic spectra taking into account 
the homogeneous and heterogeneous interactions of the four levels. Column densities
were derived from fits to the \tppzb absorption band whose oscillator strength is well established.
These are the first reported measurements for \ctdp.  For \cdsp, our results are consistent with the larger 
values obtained in the most recent laboratory and astronomical studies. 
   }
\authorrunning{M. Eidelsberg et al.}
\titlerunning{Oscillator Strengths of CO Rydberg transitions ...}
\maketitle
%

\section {Introduction}
The column density of CO and its isotopomers in diffuse interstellar clouds is derived from vacuum ultraviolet (VUV)
absorption bands seen in spectra acquired with the {\it Hubble Space Telescope (HST)} and the {\it Far Ultraviolet Spectroscopic 
Explorer (FUSE)}. Such absorption is the only technique providing direct access to the column density. However, the 
accuracy of the results obtained is entirely dependent on that of the band and line oscillator strengths used in 
the data reduction. Over the years, data on oscillator strengths provided by laboratory measurements have improved 
and many discrepancies have been resolved. In particular, a set of recommended values for the A-X band oscillator 
strengths has been proposed on the basis of converging experimental and theoretical results (Eidelsberg et al. 1999). 
These bands cover the 1300-1600 \AAb region and six orders of magnitude in oscillator strength. In the same wavelength 
range, the triplet-singlet intersystem bands are typically 2 to 6 orders of magnitude weaker and can be used for dense 
clouds where the A-X bands are optically thick. The large discrepancies that initially appeared between observed 
and predicted oscillator strengths for these transitions (Federman et al. 1994; Sheffer, Federman, Lambert 2002) 
have been resolved thanks to a new theoretical model that allows a reliable calculation of band and line oscillator 
strengths (Rostas et al. 2000; Eidelsberg and Rostas 2003).

      The availability of FUSE has extended the accessible wavelength range to the 905-1187 \AAb region where CO 
absorbs through transitions from the ground state to Rydberg states such as B$^1$$\Sigma$$^+$, C$^1$$\Sigma$$^+$,  
E$^1$$\Pi$, etc... The oscillator strengths for these transitions have often been measured in the laboratory but 
significant differences still exist between the sets of results. The available laboratory data have been reviewed 
by Federman et al. (2001) and Sheffer et al. (2003) and compared by the latter to their recent FUSE results. 
These comparisons clearly show that the available results for bands below 1075 \AAb still show too much scatter 
to provide a consensus which could lead to a set of recommended f-values for use in VUV measurements and in modelling 
CO photochemistry in astronomical environments.

      In the light of this situation, the availability of the new SU5 high resolution beam line on the Super-ACO synchrotron 
at Orsay has allowed us to undertake a systematic study of the absorption cross sections for Rydberg bands starting 
with the \tppxz  band at 1076 \AAb and up to the W-X(3-0) band at 925 \AA.  
Taking advantage of the full rotational resolution seen in most cases by the 30 \mAAb (even 20 \mAAb in the best cases) 
instrumental width, we used the simulation-fitting technique to analyse the spectra. This technique generally can 
account for overlapping of adjacent bands and for effects of line saturation which may appear in strong and sharp features.

      We present here a sample of the results obtained for three isotopomers,  \cdsp, \cts and \ctdp.  
The focus is on a spectral region
where three or four bands associated with the Rydberg levels \qpsz, \qppzb and \tdpub
[formerly K(0), L(0) and L'(1) and with the \esspb level of mixed Rydberg-valence type 
partially overlap at room temperature.
This region necessitates the use of the simulation-fitting 
technique to determine individual band oscillator strengths. The simulation of these bands is made more difficult 
by the fact that the corresponding upper states are mixed by relatively strong interactions (Sekine \& Hirose 1993). 
Therefore, they are not pure $\Sigma$ or $\Pi$ states, and the degree of mixing has to be determined for each value 
of the rotational quantum number J in order to determine the intensity of each line. Furthermore, due to isotopic 
level shifts, the interactions are different in the various isotopomers, leading to different band shapes and 
intensities. This differs from the case of isolated bands where the band intensity is not expected to vary from one 
isotopomer to the next.  In fact, the concept of band oscillator strength loses its meaning in such cases. Only the 
line intensities are well defined.  Indeed, as shown in Rostas et al. (2000), the band oscillator strengths even become
temperature dependent because the mixing coefficients vary with J and the contribution of different J levels to a given 
band changes with temperature.

      The paper is organized in the following way.  We describe next the experimental setup for our 
synchrotron-based measurements. This is followed by the data analysis in Sect. 3 including a brief description of the 
model describing the mixing of the upper states used to simulate the observed spectra. Sect. 4 presents our results 
and a discussion of them and Sect. 5 our conclusions.  

\section {Experimental setup}

The experiments were conducted on the Super-ACO synchrotron ring in Orsay, using the high spectral resolution undulator 
based SU5 beam line. The VUV spectrometer at this beam line is equipped with a 6.65 m normal incidence concave 
grating in an off-plane Eagle mounting.
Detailed characteristics of the SU5 line and specifications are presented in Nahon et 
al. (2001a, b). The main features relevant to the wavelength range (900-1080 \AA) studied in this work were the use 
of a SiC grating (2400 lines/mm$^{-1}$, blazed at 13 eV) in first order, providing a theoretical resolving power of 
$\sim$75000 at 12 eV. Entrance and exit slits of the spectrometer were set either at 35 or 25 $\mu$m, giving 
an instrumental width of 30 or 20 \mAA, respectively. The undulator current was adjusted to optimize the photon flux 
for each CO band scanned. The undulator was working in the horizontal polarization mode. We verified that the gas 
filter present on the beam line, used to reject harmonic generation in the undulator, was not necessary in our case. 
The rotation of the grating was chosen to provide a scanning step of 7 \mAAb (about one fourth of the instrumental width). 
Due to software problems, the automatic translation of the grating was disabled. The grating was positioned in the 
middle of the range corresponding to each band scanned.  Instrumental widths were measured using argon lines at 
1048.22 and 894.31 \AAb as standards. These values were consistent with those deduced from the band-fitting procedure 
described below.

      The VUV light beam emerging from the spectrometer entered a windowless cell system. Several stages of differential 
pumping were used to maintain the ultra-high vacuum inside the storage ring and the spectrometer. The 30 cm long stage 
immediately upstream from the gas cell was bounded at each end by a 1.6 mm diameter aperture. The gas cell was adjustable 
in length (usually 5.4 cm, but 3.5 cm for some bands). High purity \cds gas (Alphagaz, 99.997\%) or \cts (Eurisotop, 
99.1\% $^{13}$C, 99.95\% $^{16}$O) or \ctd (Isotech 98.8\% $^{13}$C, 94.9\% $^{18}$O) continuously flowed through 
the cell into the differential pumping section. The pressure in this section, separated from the monochromator 
chamber by a 1.6 mm aperture, was about 1000 times lower than in the gas cell. The flow was controlled by a needle 
valve that, once adjusted, maintains a constant pressure in the cell for long periods of time. A 10 Torr full scale 
capacitor gauge measured the gas pressure in the cell which was generally in the 
1-10 mTorr range. Using the 10 mTorr scale, the highest sensitivity for the gauge, leads to measurement uncertainty
of $\sim$0.05 mTorr. Great care was also taken to compensate for 
zero shifts, through periodic checking of the zero pressure reading. The pressure stability was also monitored 
by a cold cathode gauge in the differential pumping section where the pressure is in the 10$^{-6}$ Torr range. 
The flux of synchrotron radiation emerging from the gas cell was measured by a photomultiplier (EMI 9558QB working 
at -1250 V) by means of fluorescence from a fresh sodium salicylate coating deposited on the glass end-window closing 
the cell. The PMT anode current was converted to voltage and digitized by a charge amplifier (Keithley type 6485 
picoammeter) and then recorded via a GPIB bus on a computer.

       To minimize possible errors in the column density arising from weak absorption in 
the pumping sections and from possible impurities
contributing to the total pressure, we used as a standard reference the 
\tppzb band at 1076 \AA, whose oscillator strength is well characterized (Federman et al. 2001). For a given pressure, 
scans of this band were systematically obtained before and after the scans of the band of interest. The column 
density derived from these spectra in most cases was in good agreement with that provided by the pressure gauge. 
Spectral scans of the 967-972 \AAb region were obtained at 4 different pressures between 2 and 9 mTorr. 
Several scans were acquired at each pressure as were 2 or 3 scans of the \tppzb reference band. 
During the course of a set of measurements the pressure was found to vary by 5 to 10\%, this 
leading to corresponding uncertainties in the measurements of the absolute oscillator strength. The scanning step 
size was usually set at 7 \mAAb for the 35 $\mu$m slit (or 5 \mAAb for the 25 $\mu$m slit). The dark current 
of the PMT was measured before and after each scan by closing a mechanical shutter on the beam line. 
To take into account the temporal decrease of the beam flux, continuum level (I$_0$) scans were 
periodically interspersed within the molecular band records, using the same wavelength range at the same scanning 
rate as for the adjacent scans but with an empty cell.

\section{Data analysis}
\subsection{The simulation-fitting procedure}
      After subtraction of the dark current, the recorded intensities are transformed into optical depths using the 
Beer-Lambert relation, $\tau$($\lambda$)=ln[I$_0$($\lambda$)/I($\lambda$)], where I$_0$($\lambda$) and
I($\lambda$) are the incident and transmitted intensities 
respectively. The optical depth $\tau$($\lambda$) is related to the absorption cross section $\sigma$($\lambda$), 
the column density nl, and the oscillator strength f($\lambda$) by the relation 
$\tau$($\lambda$)=nl $\sigma$($\lambda$)=8.85 10$^{-21}$ $\lambda$$^2$ nlf($\lambda$),
where $\lambda$ is in \AA, n in cm$^{-3}$, l in cm and $\sigma$ in cm$^{2}$. Changes in the incident 
intensity I$_0$, experimentally measured over the course of any one scan, were less than 5\%. I$_0$ was assumed to vary 
linearly with time (i.e., wavelength) and estimated by a linear interpolation between the intensities recorded at 
the beginning and the end of the scan. 

      The corresponding simulated spectrum $\tau$($\lambda$) was calculated, as described below, on the basis of the model 
established recently by Eidelsberg et al. (2004). 
For each rotational line, the calculated wavelength 
and the proper H$\ddot{o}$nl-London factor (i.e., rotational line strength) were introduced. Each line was convolved with 
a Voigt profile comprising a Gaussian component corresponding to the room temperature Doppler width and, when 
appropriate, a Lorentzian component corresponding to the predissociation lifetime of the upper level of the transition. 

      The optically thin simulated spectrum thus calculated is transformed into an absorption spectrum using the 
Beer-Lambert law. This spectrum is then convolved with the instrument profile (taken to be Gaussian). As discussed 
in Eidelsberg et al. (1999) this procedure reproduces the line saturation effects that appear in the experimental 
spectra when the instrumental width is larger than the spectral line width.

      The simulated spectrum is then adjusted to match the experimental spectrum in a non-linear least-squares 
fitting procedure in which the band oscillator strengths and the instrumental width are free parameters. When 
predissociation is present, the Lorentzian component can also be left free to adjust in the fit and the resulting 
value used to determine the predissociation lifetime. 

      The procedure just described was executed using two independent codes previously applied to derive CO 
oscillator strengths, one by Federman et al. (1997) for A-X transitions and for the B-X, C-X, and E-X 
Rydberg transitions, the other by Eidelsberg et al. (1999) for A-X bands and by Rostas et al. (2000) for 
intersystem transitions.  The agreement in results from the two codes was excellent, typically within 2\%.

 \begin{figure}
\resizebox{\hsize}{!}{\includegraphics{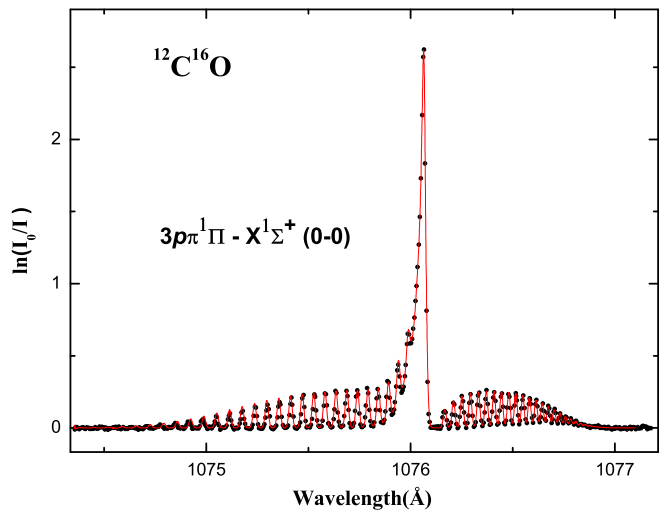}}
\caption{Fit of the simulated spectrum of the \tppxz to the experimental one for a 
column density of 9.75 10$^{14}$ cm$^{-2}$ and an instrumental width of 19.6~($\pm$0.6) \mAA.}
\end{figure}
      
The procedure was applied first to the \tppzb band used as a standard in order to determine 
reliable experimental column densities notwithstanding variations in the pressure gradients along the line of sight 
in the vacuum vessel. The 3{\it p$\pi$}(v'=0) state is far from others and can be considered as practically free of 
perturbations. Thus, the oscillator strength of the band can be assumed to be the same for CO and its isotopomers 
\cts and \ctd (see Stark et al. 1992). The synthetic spectra of this band were calculated using the H$\ddot{o}$nl-London 
factors for a $^1$$\Pi$-$^1$$\Sigma$$^+$ transition (Herzberg 1950) and a new set of unpublished line wavelengths 
extending to J = 30 (Eidelsberg et al., unpublished data) obtained through a fit of the 
available experimental data concerning the 
E-X, E-A, E-B, and E-C transitions for CO and its isotopomers. The line shape included a Doppler component 
of 2.5 \mAAb and a Lorentzian component of 0.067 \mAAb for CO, 0.049 \mAAb for \ctsp, and 0.047 \mAAb for \ctd (Cacciani et 
al. 1988). The Lorentz part of the line width is too small in this case to be determined by the least-squares fit. 
The rotational temperature was set at 295 K. The column density was obtained by setting the \tppzb band oscillator 
strength  to the value determined by Federman et al. (2001) and by fitting the synthetic \tppzb spectrum to the experimental 
data. Initially, the instrumental width was allowed to vary. For a given setting of the slits, the inferred 
width was found to be constant and close to that obtained from the profile of the argon line at 1048 \AA. 
The instrumental width was then kept fixed for a given set of measurements as long as the slits were not changed. 
Fits were made on the R, P, and Q branches separately and the individual column densities were averaged. A typical 
fit is shown in Fig. 1 for the \tppzb band in \cdsp. The absorption cell length was 3.52 cm and the measured pressure 
8.4 mTorr. The instrumental width obtained from the fit was 19.6$\pm$0.6 \mAAb and the column density was 
(9.75$\pm$1.00)10$^{14}$ cm$^{-2}$.

\subsection{Calculation of simulated spectra for the perturbed Rydberg transitions}
      The simulated spectra are calculated taking into account the interactions between the upper states of the
four bands. The line positions and the interaction parameters between the 
states have been determined recently for four isotopomers by Eidelsberg et al. (2004). Proper use of 
this model allows us to calculate realistic spectra for the bands under study.

	The deperturbation analysis by Eidelsberg et al. (2004) shows the upper states of the bands considered 
here to result from the interaction of the \xpzqp complex with the \xputdp component and
with the {\it E(6)} level which is of strongly mixed Rydberg-valence type.
The deperturbed molecular parameters and the interaction energies are given in Table VII of 
Eidelsberg et al. (2004). These values are introduced in the Hamiltonian energy matrix given in the 
same reference. Diagonalizing this matrix for each value of J provides the energies of the resulting observed 
rotational levels T$_i$(J) and the mixing coefficients \cij that describe the contribution of the deperturbed
state j to the mixed wavefunction of state i. As an example the mixing coefficients 
\cij for \qpszb obtained from this procedure are displayed in Figs. 2, 3, and 4. 
These figures illustrate the mixing of the \qpszb  with the {\it e}  parity levels of the \spib perturbers (l~uncoupling). 
Similar figures for the {\it e} and {\it f} parity levels  of the 1\spi, 2\spib and 3\spib levels which 
arise from much stronger homogeneous interactions can also be traced.

\begin{figure}
\resizebox{\hsize}{!}{\includegraphics{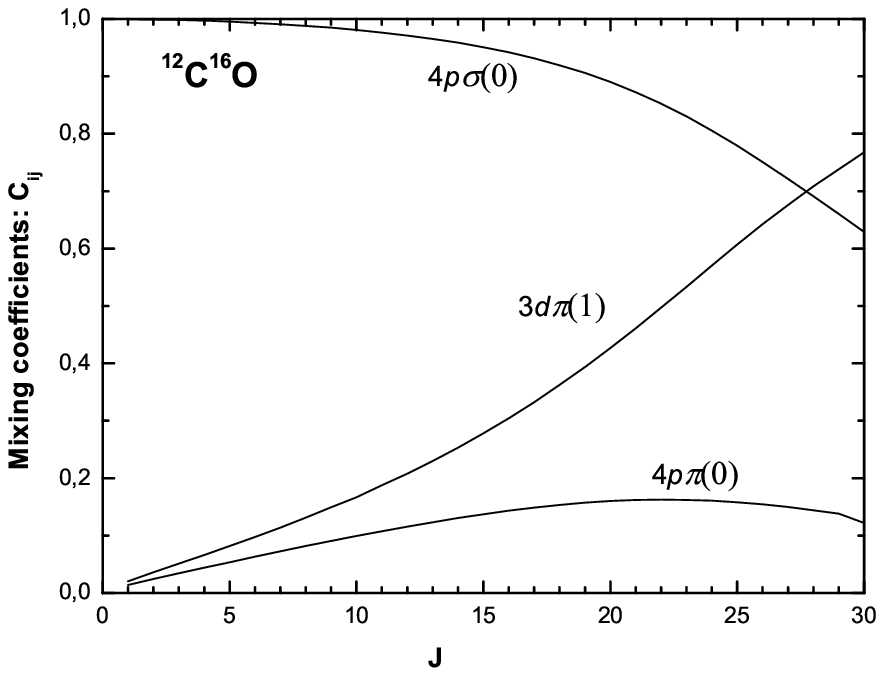}}
\caption{Mixing coefficient of the \qpszb level  in  \cdsp.}
\end{figure}

\begin{figure}
\resizebox{\hsize}{!}{\includegraphics{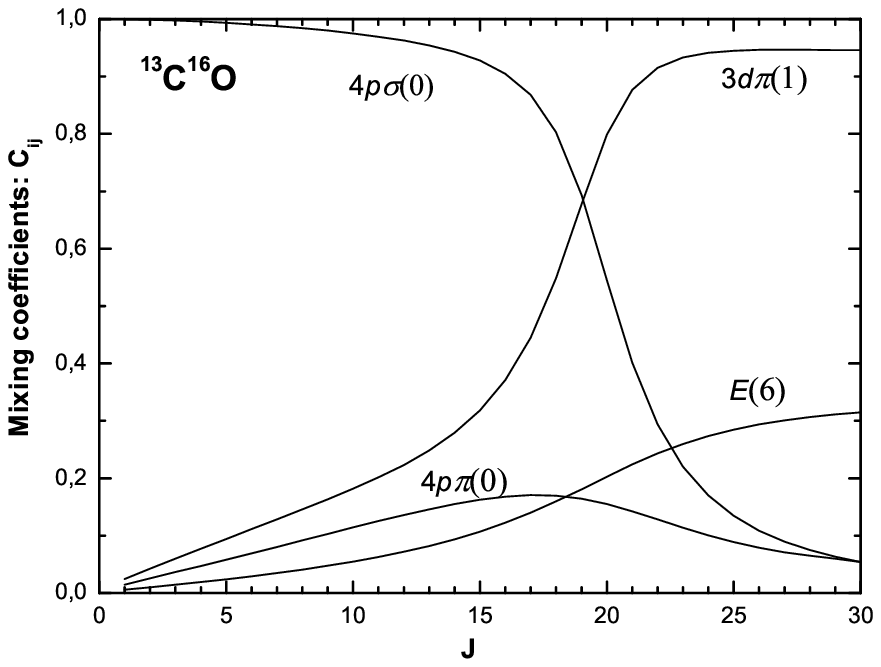}}
\caption{Mixing coefficient of the \qpszb level  in  \ctsp.}
\end{figure}

\begin{figure}
\resizebox{\hsize}{!}{\includegraphics{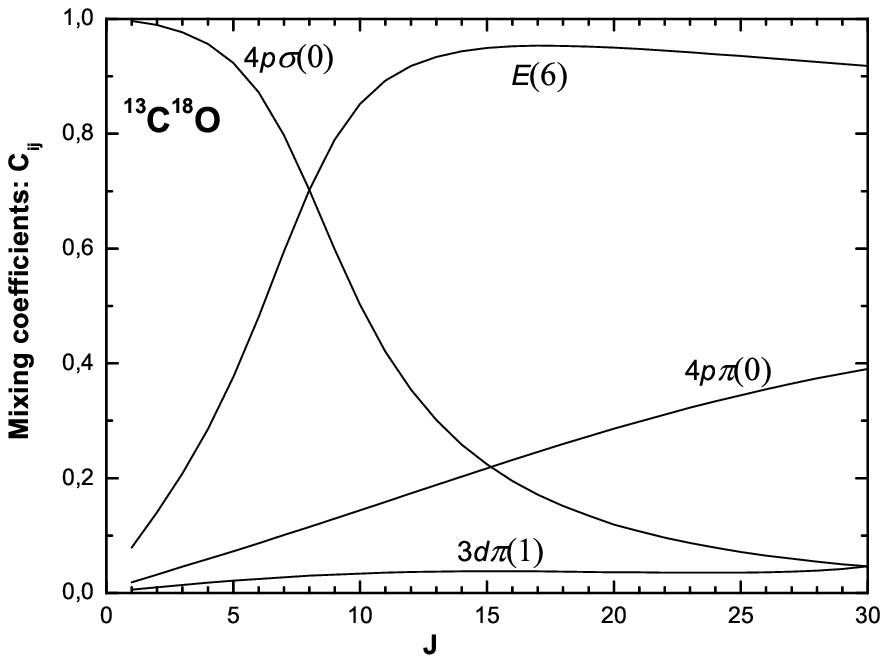}}
\caption{Mixing coefficient of the \qpszb  level in  \ctdp.}
\end{figure}
     
      Following the diagonalization of the energy matrix, the line wavenumbers are calculated as the difference 
between the energies of the relevant upper and lower levels. The line strengths are obtained as described in 
Lefebvre-Brion \& Field (1986; Eqs. 5.3.29a,b)
which give the perturbed transition amplitudes for transitions between a pair of mutually interacting
\sspb and \spib upper states and an unperturbed \sspb  lower state. The expressions for the transition amplitudes 
include the mixing coefficient, the H$\ddot{o}$nl-London  factor and one perpendicular and one parallel transition moment. 
Extending these equations to the case of four interacting upper states 
(three $\Pi$ and one $\Sigma$) leads to the addition of two more transition moments, both of perpendicular type.
The Doppler width of CO absorption lines at room temperature, at 970 \AAb is 2.3 \mAA, which is less 
than the measured instrumental width of 20 \mAA. For the {\it L(0)} and {\it K(0)} bands, the Lorentz part of the line width 
is even smaller. They were set to the values of Drabbels et al. (1993), i.e.: 1.1 and 0.09 \mAA, respectively, for 
\cdsp. For \cts and \ctdp, the values given by Cacciani et al. (2002) for {\it L(0)} were adopted (respectively, 
0.7 and 1.7 \mAA); as for the {\it K(0)} and {\it E(6)} bands for these isotopomers, 1.7 \mAAb was used, corresponding to the 
lower limit of 3.6 10$^{-10}$ s for the lifetime given by Ubachs et al. (1994). The {\it L'(1)} band however, is diffuse 
and the Lorentz component of the width is not negligible. A value of 15$\pm$5 \mAAb was found through our fits and 
was used for all three isopotomers.

 The calculated spectra are fitted to the experimental ones using a nonlinear least-squares method. 
This procedure yields integrated cross sections for each 
of the bands involved, independently of their overlap. A test of the quality of the process is provided by the 
fact that the sum of the integrated cross sections of the three bands should be the same for the three isotopomers.

\section{Results and discussion}

      The band system in the 967-972 \AAb range was scanned for the three isotopomers, \cdsp, \cts and \ctdp, 
at several pressures in the 1 to 10 mTorr range. Figs. 5-7 illustrate typical results for each of the three 
isotopomers. In these figures and from now on the observed mixed levels of \spib symetry are referred following the 
notation of Eidelsberg et al. (2004), as 1\spi, 2\spib and 3\spi, in order of increasing energy. 
In each figure,  trace (a) represents the calculated spectra of the individual bands, (b) the sum of these spectra, 
c) the experimental spectrum and (d) the residuals obs-calc. The model reproduces the 
evolution of the spectrum when going from one isotopomer to the next very well. 

\begin{figure*}
\resizebox{\hsize}{!}{\includegraphics{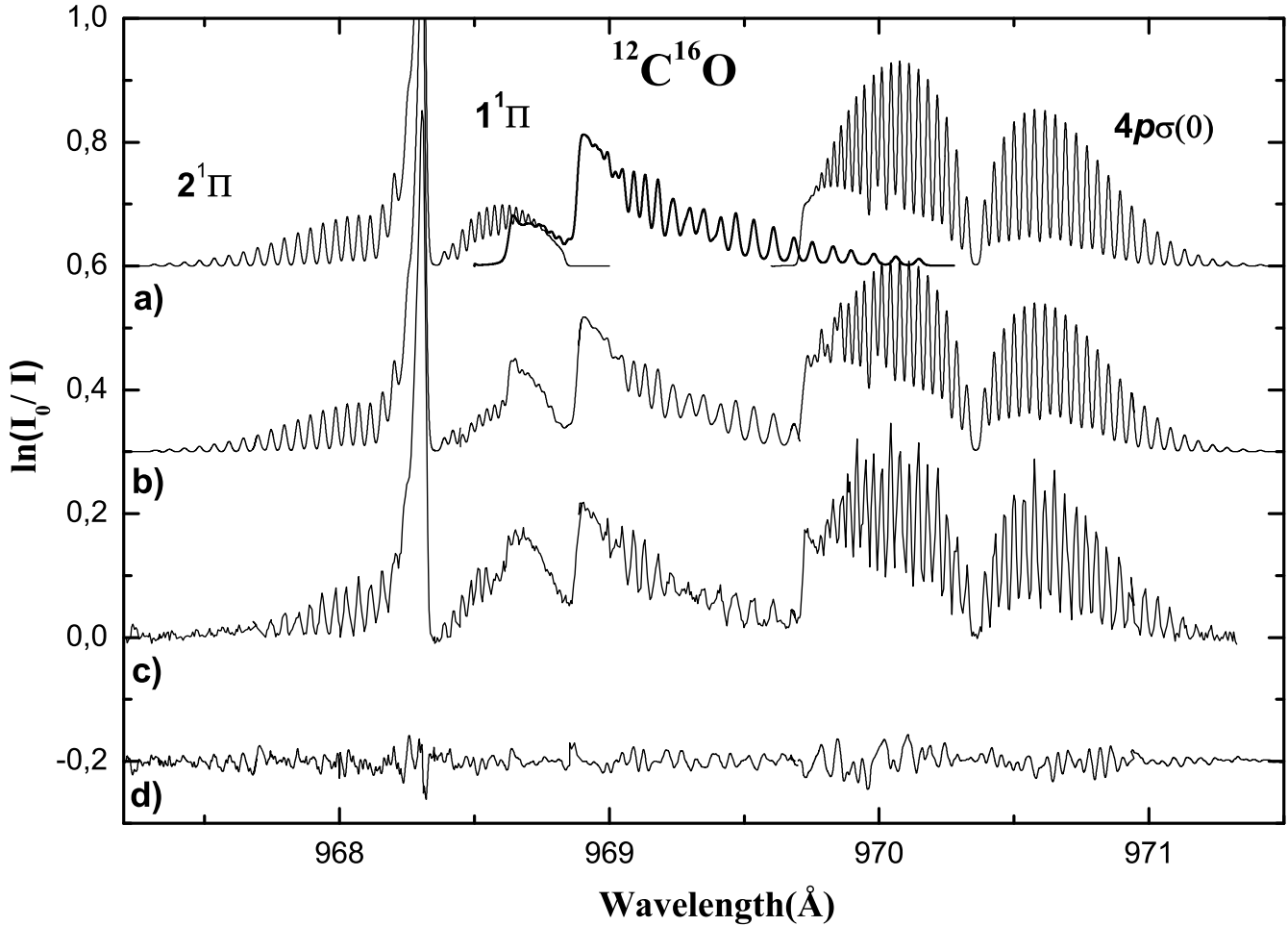}}
\caption{ 
a) Simulated spectra of the 3 bands 1\spi-X, 2\spi-X and \qpsz-X in
  \cds taking into account the l~uncoupling interaction between the two 1\spib levels with the \xsmix level.
          b) Fit of the 3 bands together.
          c) Experimental spectra for a column density of 9.75 10$^{14}$cm$^{-2}$ and an experimental width of  20 \mAA.
          d) Quality of the fit.
}
\end{figure*}

\begin{figure*}
\resizebox{\hsize}{!}{\includegraphics{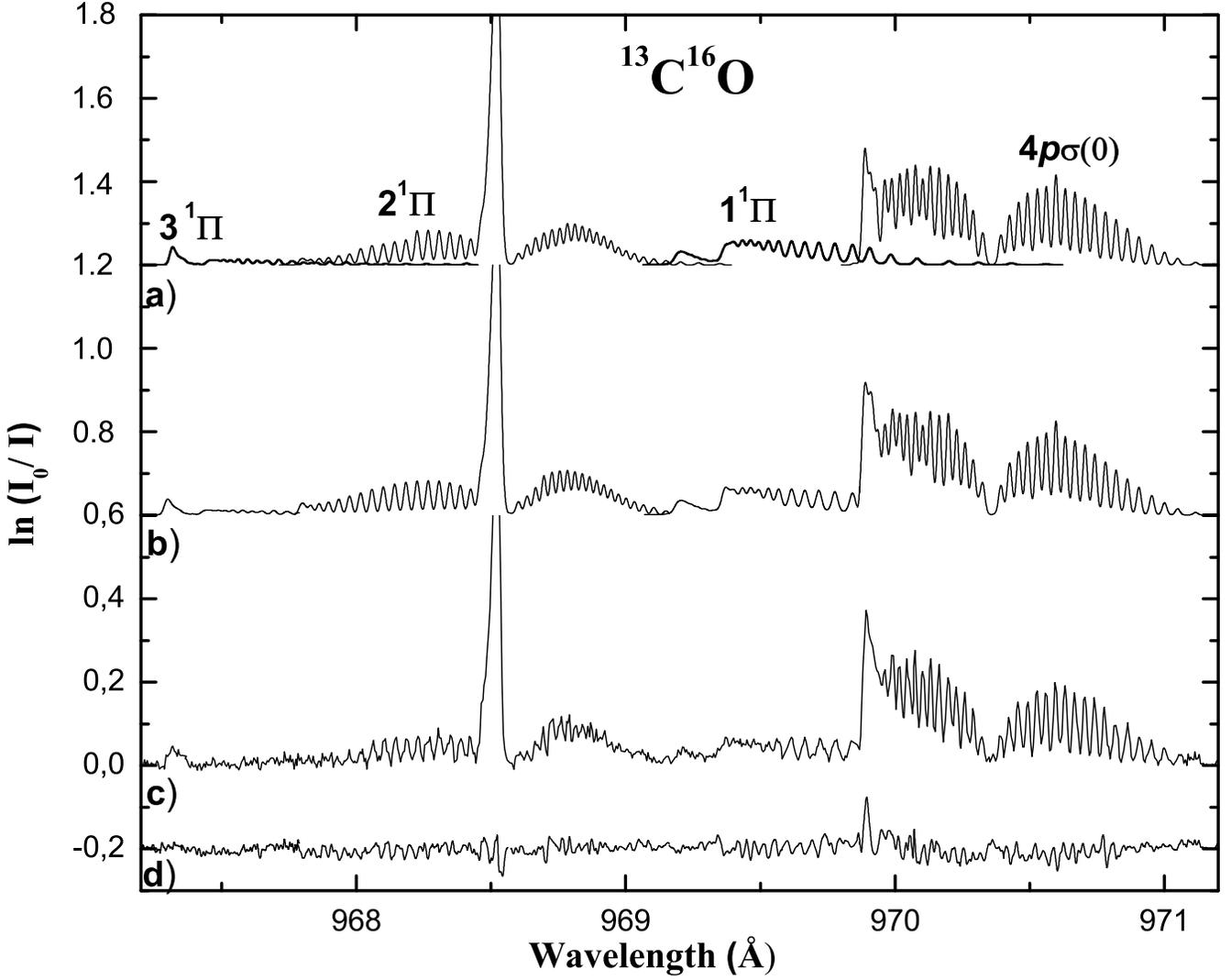}}
\caption{ a) Simulated spectra of the 4 bands 3\spi-X, 2\spi-X, 1\spi-X, and \qpsz-X in
in \cts taking into account the l~uncoupling interaction between 
the \qpszb level with the \qppzb level, and the homogeneous interactions of the \qppzb and \tdpub levels 
with the {\it E(6)} perturber.
          b) Fit of the 4 bands together.
          c) Experimental spectra for a column density of 1.8 10$^{14}$cm$^{-2}$ and an experimental width of  20 \mAA.
          d) Quality of the fit.}
\end{figure*}

\begin{figure*}
\resizebox{\hsize}{!}{\includegraphics{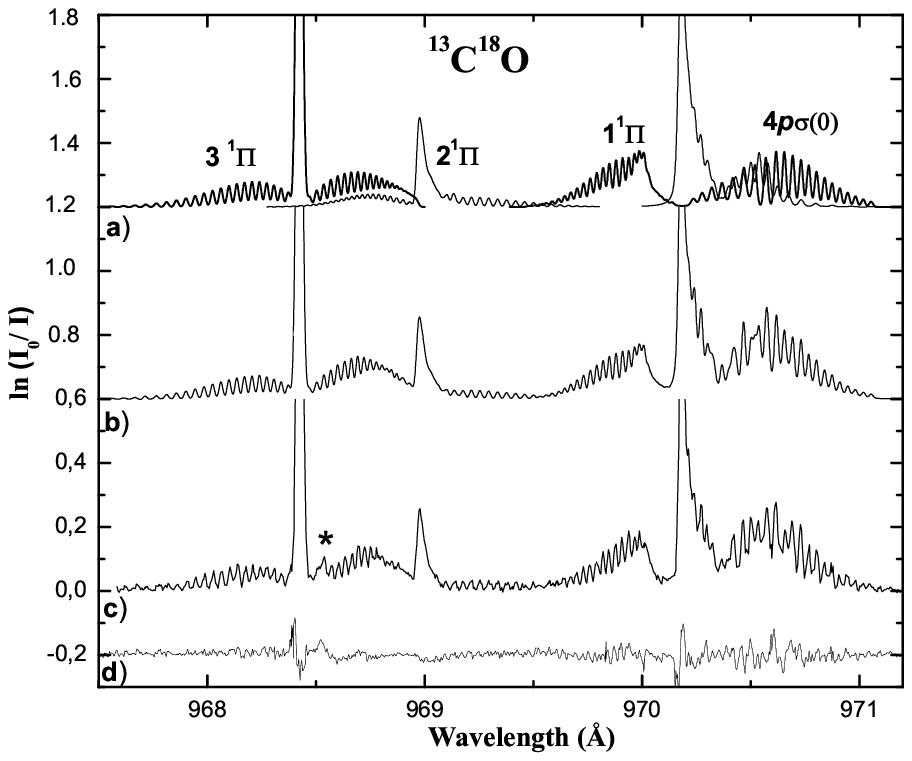}}
\caption{ a) Simulated spectra of the 4 bands 3\spi-X, 2\spi-X, 1\spi-X, and \qpsz-X in
in \ctd taking into account the l~uncoupling interaction between 
the \qpszb level with the \qppzb level, and the homogeneous interactions of the \qppzb and \tdpub levels 
with the {\it E(6)} perturber.
          b) Fit of the 4 bands together.
          c) Experimental spectra for a column density of 2.6 10$^{14}$cm$^{-2}$ and an experimental width of  20 \mAA.
          d) Quality of the fit.
(The asterisk $\ast$ indicates a contamination of the P branch by the Q branch of the same transition in \ctsp.)
}
\end{figure*}

\renewcommand{\arraystretch}{0.8}
\tabcolsep=0.12cm

\begin{table*}
\begin{center}
\caption[ ]{Oscillator strengths for $^{12}$C$^{16}$O bands between 967 and 972
{\AA } (f-values $\times$ 10$^{-3}$)}
\begin{tabular}{ccccclccc}
\hline
\noalign{\smallskip}
 Band & Band Origin & This work & Letzelter et al. & Stark et al.   & Stark et al. & Stark et al. & Yoshino et al. & Sheffer et al. \\
  & & & 1987$^{\ast}$ & 1991$^{\ast}$ &   1993$^{\ast}$ & 1993$^{\ast}$ & 1995$^{\ast}$ & 2003 \\
 (upper level) & $\lambda$$_{0}$ ({\AA }) & \bf{295 K} & \bf{295 K} & \bf{295 K}   & \bf{295 K} & \bf{20 K} & \bf{20 K} & \bf{ISM} \\
\noalign{\smallskip}
\hline
\noalign{\smallskip} 
 \qpszb$^{a}$ & 970.357 & 34.1$\pm$3.4 & 21.0$\pm$2.1 & 13.4$\pm$1.3 &  26.0$\pm$2.6  & 26.8$\pm$5.3 & 33.5$\pm$5.0 & 31$\pm$4\\
  &  &  &  &  &   &   & 34.8$^{b}$ & 35.0$^{c}$ \\
 1\spib$^{a}$ & 968.889 & 12.0$\pm$1.2 & 12.4$\pm$1.2 & 8.5$\pm$0.8 & $\rceil$  & 11.4$\pm$2.3 & $\dots$ & 10.1$\pm$1.1\\
 &  &  &  &  &   22.4$\pm$2.4 &  &  & 11.19$^{c}$ \\
 2\spib$^{a}$ & 968.300 & 14.5$\pm$1.5 & 7.7$\pm$0.8 & $>$8.9  & $\rfloor$  & 7.5$\pm$0.8 & 11.9$\pm$1.8 & $\dots$\\
 &  &  &  &  &  &  & 13.85$^{b}$ &  \\
 \noalign{\smallskip} 
\hline
\noalign{\smallskip} 
\qpszb + 1\spib + 2\spib & $\dots$ & 60.6$\pm$6.1 & 41.1 & $>$30.8  &  48.4 & 45.7 & $\dots$ & $\dots$\\
\noalign{\smallskip}
\hline
\end{tabular}
\end{center}
* For comparison, the integrated absorption cross sections given in these references have been transformed into 
oscillator strengths through the relation f = (113 $\times$ $\int$$\sigma$($\lambda$) d$\lambda$)/$\lambda
$$^{2}$, where the cross section $\sigma$($\lambda$) is in 10$^{-18}$ cm$^{2}$ and $\lambda$ is in {\AA }.\\
$^{a}$ \qpsz, 1\spib and 2\spib refer to {\it K}, {\it L} and {\it L'}, respectively, in the old notation.\\
$^{b}$ Our calculated value at 20 K.\\
$^{c}$ Our calculated value at 4 K.
\end{table*}

\begin{table*}
\begin{center}
\caption[ ]{Oscillator strengths for $^{13}$C$^{16}$O bands between 967 and 972
{\AA } (f-values $\times$ 10$^{-3}$)}
\begin{tabular}{cccc}
\hline
\noalign{\smallskip}
 Band & Band Origin & This work & Eidelsberg \& Rostas \\
 (upper level) & $\lambda$$_{0}$ ({\AA }) &  & 1990 \\
\noalign{\smallskip}
\hline
\noalign{\smallskip} 
 \qpszb$^{a}$ & 970.356 & 32.64$\pm$3.26 & 18.30$\pm$1.83   \\
 1\spib$^{a}$ & 969.359 & 6.99$\pm$0.70 & 11.42$\pm$1.14   \\
 2\spib$^{a}$ & 968.532 & 20.72$\pm$2.07 & 12.77$\pm$1.28   \\
 3\spib$^{a}$& 967.426  &  1.20$\pm$0.12  &  \\
\noalign{\smallskip} 
\hline
\noalign{\smallskip} 
 \qpszb+ 1\spib+ 2\spib+ 3\spib & $\dots$ & 61.55$\pm$6.16  & 42.49$\pm$4.25 \\
\noalign{\smallskip}
\hline
\end{tabular}
\end{center}
$^{a}$ \qpsz, 1\spib and 2\spib refer to {\it K}, {\it L'} and {\it L}, respectively, in the old notation,
while 3\spib corresponds to the newly characterized {\it E(6)}.
\end{table*}

      The fitting procedure provides, for a given spectrum, the oscillator strengths of each component 
band. The results of the present study are given in Tables 1-3 with a comparison to others that are available.  
The oscillator strength for a given band varies from one isotopomer to the next because the amount of mixing between 
bands differs. However, the total oscillator strength of the band system should not be isotope sensitive as it is 
the sum of those of the deperturbed bands which themselves are not sensitive to isotope effects. This expectation is 
verified within about 10\% and provides a good test of the overall quality of our results which depends both on 
the model used and on the experimental procedure.

\begin{table*}
\begin{center}
\caption[ ]{Oscillator strengths for $^{13}$C$^{18}$O bands between 967 and 972
{\AA } (f-values $\times$ 10$^{-3}$)}
\begin{tabular}{ccc}
\hline
\noalign{\smallskip}
 Band & Band Origin & This work  \\
 (upper level) & $\lambda$$_{0}$ ({\AA }) &   \\
\noalign{\smallskip}
\hline
\noalign{\smallskip} 
  \qpszb$^{a}$ & 970.365 &  22.32$\pm$2.23  \\
  1\spib$^{a}$ & 970.139 &   20.90$\pm$2.09 \\
  2\spib$^{a}$ & 968.964 & 5.57$\pm$0.56   \\
  3\spib$^{a}$ & 967.529 &  18.06$\pm$1.80  \\
\noalign{\smallskip} 
\hline
\noalign{\smallskip} 
\qpszb+ 1\spib+ 2\spib+ 3\spib & $\dots$   &  66.85$\pm$6.69  \\
\noalign{\smallskip}
\hline
\end{tabular}
\end{center}
$^{a}$ \qpsz, 2\spib and 3\spib refer to {\it K}, {\it L'} and {\it L}, respectively, in the old notation,
while 1\spib corresponds to the newly characterized {\it E(6)}.
\end{table*}

      The present results are also compared to previous ones in Tables 1-3. The comparison is straightforward 
for other data recorded at room temperature. However, the data of Stark et al. (1993) and Yoshino et al. (1995) 
obtained at 20 K and the 4 K interstellar (ISM) data of Sheffer et al. (2003) cannot, in principle, be compared 
directly. Instead, we recalculated synthetic spectra for 20 and 4 K and obtained band oscillator strengths for 
comparison with the values shown in Table 1. The agreement between our inferred oscillator strengths and the 
recent determinations (Yoshino et al. 1995; Sheffer et al. 2003) is excellent, being within the quoted 1-$\sigma$ 
empirical uncertainties.  The band values previously obtained experimentally at room temperature without the 
help of the fitting technique are somewhat arbitrary due to the overlaps. Therefore, only the oscillator strengths 
for the total band system are comparable. Our room temperature results tend to be about 50\% larger than the 
previous ones, for both \cds and \ctsp. This could be attributed to possible optical depth effects which 
might have affected earlier results (see also Federman et al. 2001).

\section{Conclusions}

      We carried out absorption experiments using synchrotron radiation to measure oscillator strengths for 
several overlapping Rydberg transitions in \cdsp, \cts and \ctdp.  In this paper particular attention 
was devoted to the {\it L(0)}, {\it L'(1)}, and {\it K(0)} bands of all three isotopomers as well as the 
{\it E(6)} band in the two rarer forms of CO.  The necessary line strengths and mixing coefficients were 
obtained from the deperturbed molecular parameters given in Eidelsberg et al. (2004).

      As for the bands at wavelengths greater than 1075 \AA, consensus is emerging regarding oscillator 
strengths for the {\it L(0)}, {\it L'(1)}, and {\it K(0)} bands of \cdsp; our results are consistent with other recent 
determinations.  This consistency among results for \cds and the fact that the sum of oscillator strengths 
for the set of overlapping bands is independent of isotopomer give us confidence in our results on \cts and 
\ctdp.  Since absorption involving these bands plays an important role in CO photodissociation in astronomical 
environments (e.g., van Dishoeck \& Black 1988), use of our larger oscillator strengths for \cds and \cts 
and our new ones for \ctd may help resolve the differences now found between model predictions and observations 
(e.g., see Sheffer et al. 2003; Federman et al. 2003).

\section{Acknowledgments}
The authors are obliged to Joelle Rostas for useful discussions and to K.P. Huber for critically reading the manuscript and for 
his suggestions. They also acknowledge the support of the LURE-Super ACO facility and the SU5 beam line team.  
This research was funded in part by the CNRS-PCMI program and NASA grant NAG5-11440.

\end{document}